# Asymmetric dynamics of edge exchange spin waves in honeycomb nanoribbons with zigzag and bearded edges boundaries

D. Ghader[(1)] and A. Khater[(2,3)]

[1] College of Engineering and Technology, American University of the Middle East, Kuwait

[2] Department of Theoretical Physics, Institute of Physics, Jan Dlugosz University in Czestochowa, Am. Armii Krajowej 13/15, Czestochowa, Poland

[3] Department of Physics, University du Maine, 72085 Le Mans, France

**We report on the theoretical prediction of asymmetric edge spin waves, propagating in opposite directions on the edges of honeycomb nanoribbons with zigzag and bearded edges boundaries. The simultaneous propagation of edge spin waves in the same direction on both edges of these nanoribbons is hence predicted to be forbidden. These asymmetric edge exchange spin waves are analogous to the nonreciprocal surface spin waves reported in magnetic thin films. Their existence is related to the nontrivial symmetry underlying the nanoribbons under study. The edge and discretized bulk exchange spin waves are calculated in the long wavelength part of the Brillouin zone using the classical spin wave theory approach and appropriate boundary conditions. In the absence of an external magnetic field in our study, the asymmetric edge spin waves propagate with equal frequencies and opposite directions, since the energy dispersion relation is independent of the sign of the wavevector components in the long wavelength part of the Brillouin zone. The edge spin waves are characterized by linear dispersion relations for magnetically isotropic nanoribbons. Introducing magnetic anisotropy in the calculation significantly enhances the energy gap between the edge and bulk spin waves in both types of nanoribbons. Based on our calculation, the large energy gap allows separate excitation of bulk and edge modes as their energies are no more overlapping.**

*Introduction*

The continuum classical field theory with appropriate boundary conditions is widely adopted in the study of boundary magnetic excitations in thin films [1-12] and layered structures [13-18]. Dipolar spin waves in the classical field theory are subjected to the Maxwell boundary conditions. Exchange spin waves boundary conditions can be obtained by integrating the equations of motion over a small volume that contains the boundary [19]. Another widely adopted approach for exchange surface spin waves is to require that bulk and edge spins oscillate at the same frequency for a given spin wave eigenmode [8-10]. This requires boundary spins to satisfy the bulk equations of motion and boundary equations can be derived accordingly. Bulk equations of motion are usually derived from the Bloch's equations of motion, assuming smoothly varying exchange or dipolar magnetic fields in the continuum limit. Allowed spin waves are then determined by solving the bulk equations of motion consistently with the boundary equations.

Classical field theory results for spin dynamics on thin and ultrathin magnetic films proved fundamental effects on both bulk and surface spin wave modes induced by the film's symmetry, magnetic order, surface structure and thickness. One fascinating fact in the field of thin film magnetism is the existence of nonreciprocal localization of surface spin waves in magnetic films with special symmetry and magnetic ordering. In these magnetic films, surface spin waves confined to opposite surfaces travel in opposite directions and are characterized by different frequencies in the presence of an external magnetic field. These films do not allow the simultaneous propagation of surface spin waves in the same direction on both boundaries. A comprehensive discussion on the symmetries giving rise to nonreciprocal surface spin waves in magnetic films can be found in [9]. Nonreciprocal spin waves have received intensive theoretical and experimental attention in view of their importance for technological applications [20-24].

The important recent discovery of magnetic properties in Dirac materials attracted substantial interest in their magnetic excitations [25-37]. The novel characteristics for magnetic excitations in Dirac materials is believed to open important new opportunities in the field of magnonics. Bounded 2D or quasi 2D Dirac materials present bulk and edge spin waves, in analogy with bulk and surface spin waves in magnetic thin films. Edge spin waves on semi-infinite and nanoribbon honeycomb monolayers with zigzag and armchair boundaries have been theoretically studied using quantum spin waves approaches. The edge spin waves are found to display interesting and unconventional characteristics, notably for magnetic isotropic monolayers with zigzag edges boundaries, where Dirac edge modes are predicted.

The classical field spin wave theory is known to be equivalent to quantum spin wave approaches in the linear spin wave approximation. Despite its success in the study of surface spin waves, the classical field theory has not been systematically developed for edge spin waves in 2D materials, until our recent study of long wavelength exchange spin wave modes on nanoribbons with armchair edges boundaries [38]. Our previous study highlighted the important consequences on bulk and edge exchange spin waves induced by the finite width of the nanoribbon and the magnetic exchange anisotropy. We reported on the bulk spin wave spectrum discretization due to the finite width of the nanoribbon with armchair boundaries. To our knowledge, this was reported for the first time for honeycomb nanoribbons, as previous quantum studies neglected the effect of the nanoribbon width.

In the present work, we further develop the theory and apply it to study the bulk and edge exchange spin waves in nanoribbons with zigzag and bearded edges boundaries, in the long wavelength part of the Brillouin zone. To our knowledge, the spin wave excitations in nanoribbons with bearded edges boundaries have not been studied previously. Appropriate boundary conditions are derived taking into account the finite width of the nanoribbons. Solving the bulk equations of motion consistently with the boundary conditions yields the edge and discretized bulk exchange spin waves for both types of nanoribbons in the long wavelength part of the Brillouin zone. Unlike previous theoretical studies [25, 33, 34, 37] where the boundary conditions are derived and solved on one edge of the nanoribbon, in our approach the derived boundary equations are solved simultaneously on both edges. Edge spin waves are obtained for both types of nanoribbons, with and without magnetic anisotropy. Our results for the edge modes of the magnetically isotropic zigzag edged nanoribbon are identical to the results of existing quantum studies [25, 33], as the decay factors are found to be independent of the nanoribbon width. For the anisotropic zigzag edged nanoribbon, however, our results are significantly different. The nontrivial symmetry common for the two types of nanoribbons is found to have important consequences on the characteristics of the edge spin waves. We theoretically predict that nanoribbons with zigzag or bearded edges boundaries forbid the simultaneous propagation of edge spin waves in the same direction on opposite edges. The allowed edge spin waves hence propagate in opposite directions on the opposite edges of the nanoribbons, in analogy with the nonreciprocal surface spin waves in magnetic thin films. In the absence of an external magnetic field in our study, the edge spin waves propagate with equal energies, since the dispersion relation is independent of the sign of the wavevector components. Our study also demonstrates the important consequences induced by anisotropy on the energy gaps between the edge and propagating modes energies.

The honeycomb nanoribbons with zigzag and bearded edges boundaries are presented in figure 1. The nanoribbons are considered infinite in the x-direction, finite in the y-direction, with edges at $x = \pm d$. In terms of the honeycomb lattice constant $a$, $d$ is respectively equal to $\frac{3n+2}{2\sqrt{3}}a$ and $\frac{3n+1}{2\sqrt{3}}a$ for zigzag and bearded edges nanoribbons ($n$ is an integer). In the Néel antiferromagnetic ordering state, the spins on A (yellow) and B (blue) sublattices are conventionally assumed aligned parallel and antiparallel to the z-axis. The left and right edge spins are of A and B types respectively in the zigzag edged nanoribbons. The opposite is assumed for the nanoribbons with bearded edges boundaries.

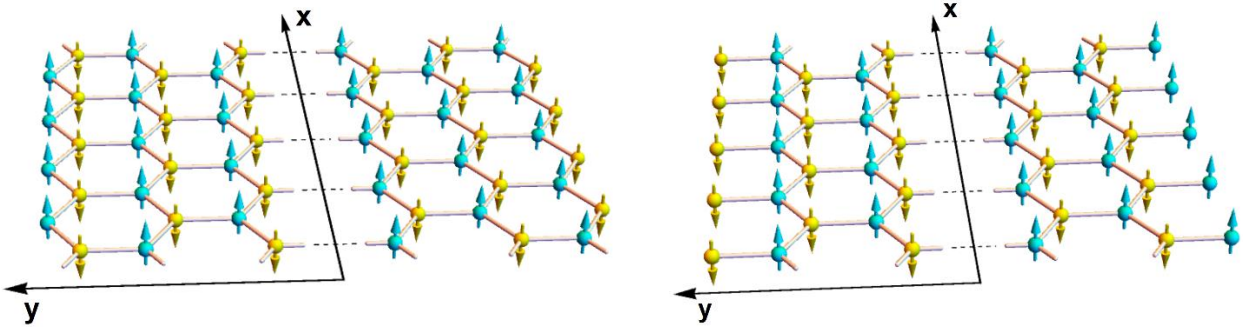

**Figure 1:** Schematic representations of a honeycomb nanoribbon with zigzag edges boundaries (left) and bearded edges boundaries (right).

## *Bulk equations of motion*

The classical field derivation for the bulk equations of motion is presented briefly. Details can be found in our previous study on nanoribbons with armchair edges boundaries [38].

The semi-classical Heisenberg Hamiltonian for exchange interaction between the spins is given as

$$\mathcal{H} = J \sum_{\langle \vec{r}, \vec{\delta} \rangle} [\vec{S}_{\|}(\vec{r}, t) \cdot \vec{S}_{\|}(\vec{r} + \vec{\delta}, t) + \gamma S_z(\vec{r}, t) S_z(\vec{r} + \vec{\delta}, t)]$$

$t$ is time, $\vec{r} = x\hat{x} + y\hat{y}$ is the position vector of a site on the honeycomb lattice, and $\vec{\delta}$ is the position vector of a nearest neighbor. $J$ is the exchange constant and $\gamma \geq 1$ is an anisotropy parameter. The vector $\vec{S}_{\|} = S_x \hat{x} + S_y \hat{y}$ represents the spin component in the plane of the honeycomb lattice.

The Bloch equations of motion for the magnetization $\vec{M}^A$ and $\vec{M}^B$ of the honeycomb sublattices are

$$\partial_t \vec{M}^A = \lambda \vec{M}^A \times \vec{H}^A \qquad (1a)$$

$$\partial_t \vec{M}^B = \lambda \vec{M}^B \times \vec{H}^B \qquad (1b)$$

with $\partial_t = \partial/\partial t$ and $\lambda$ is the gyromagnetic ratio. The vectors $\vec{H}^A$ and $\vec{H}^B$ denote the effective exchange fields acting on $\vec{M}^A$ and $\vec{M}^B$ respectively. These can be determined in terms of the magnetizations using the Heisenberg Hamiltonian. Expanding the magnetizations to second order and substituting in equations (1) yields equations of motion of the form

$$\partial_t M_x^A = -3\gamma\lambda JM\, M_y^A - \lambda JM \left(3 + \frac{a^2}{4}\Delta\right) M_y^B \qquad (2a)$$

$$\partial_t M_y^A = 3\gamma\lambda JM\, M_x^A + \lambda JM \left(3 + \frac{a^2}{4}\Delta\right) M_x^B \qquad (2b)$$

$$\partial_t M_x^B = 3\gamma\lambda JM\, M_y^B + \lambda JM \left(3 + \frac{a^2}{4}\Delta\right) M_y^A \qquad (2c)$$

$$\partial_t M_y^B = -3\gamma\lambda JM\, M_x^B - \lambda JM \left(3 + \frac{a^2}{4}\Delta\right) M_x^A \qquad (2d)$$

with $M = M_z^A = -M_z^B$ and $\Delta = \frac{\partial^2}{\partial x^2} + \frac{\partial^2}{\partial y^2} = \partial_x^2 + \partial_y^2$. With the change of variables $\mathcal{M}^A = M_x^A + iM_y^A$ and $\mathcal{M}^B = M_x^B + iM_y^B$, the equations of motion simplify to

$$-i\partial_t \mathcal{M}^A = 3\gamma\lambda JM\, \mathcal{M}^A + \lambda JM \left(3 + \frac{a^2}{4}\Delta\right) \mathcal{M}^B \qquad (3a)$$

$$i\partial_t \mathcal{M}^B = 3\gamma\lambda JM\, \mathcal{M}^B + \lambda JM \left(3 + \frac{a^2}{4}\Delta\right) \mathcal{M}^A \qquad (3b)$$

Equations (3a) and (3b) can be combined to derive the bulk spin wave equation

$$\left[\frac{1}{v^2}\partial_t^2 - \Delta + \mu^2\right] \mathcal{M}^{A/B}(\vec{r}, t) = 0 \qquad (4)$$

with $v = \sqrt{\frac{3}{2}} \lambda JMa$ and $\mu = \sqrt{\frac{6(\gamma^2 - 1)}{a^2}}$.

To solve equation (4) in the bounded system, the variables $\mathcal{M}^A$ and $\mathcal{M}^B$ are written in the general form

$$\mathcal{M}^A = A_1 e^{i(\omega t - k_x x) + q y} + A_2 e^{i(\omega t - k_x x) - q y} \tag{5a}$$

$$\mathcal{M}^B = B_1 e^{i(\omega t - k_x x) + q y} + B_2 e^{i(\omega t - k_x x) - q y} \tag{5b}$$

Compared to previous studies [25, 33, 34, 37], we here adopt a more general form for the solutions, suitable for bounded systems where both $e^{\pm q y}$ terms are physical. Here, $k_x$ is the continuous wave vector along the infinite x-direction. The real and imaginary values of $q$ correspond respectively to evanescent (edge), and propagating (bulk), spin waves in the y-direction along which the nanoribbon is finite. Substituting equation (5) in the bulk wave equation yields the dispersion relation

$$-\Omega^2 + \frac{3}{2} a^2 (k_x^2 - q^2) + 9(\gamma^2 - 1) = 0 \tag{6}$$

with the normalized frequency $\Omega$ defined as $\Omega = \frac{\omega}{\lambda J M}$.

Both bulk and edge exchange spin wave modes are determined by solving equation (6) in consistency with the boundary conditions to be derived shortly. Before proceeding to the boundary conditions, we use the bulk equations of motion (equations (3)) and determine useful relations between the coefficients $A_1$, $A_2$, $B_1$, and $B_2$. This reduces the number of independent variables which is necessary to match the number of boundary equations in the present case. Substituting equations (5) in equations (3) yields the linear equations

$$[(3\gamma - \Omega)A_1 + \zeta B_1] e^{q y} + [(3\gamma - \Omega)A_2 + \zeta B_2] e^{-q y} = 0 \tag{7a}$$

$$[\zeta A_1 + (3\gamma + \Omega)B_1] e^{q y} + [\zeta A_2 + (3\gamma + \Omega)B_2] e^{-q y} = 0 \tag{7b}$$

with $\zeta = 3 + \frac{a^2}{4}(q^2 - k_x^2)$. Equations (7) hold for any $y$ along the width of the nanoribbon. This is only possible if all coefficients of $e^{\pm q y}$ are zeros. This implies the relations

$$A_1 = -\frac{3\gamma + \Omega}{\zeta} B_1 = -\frac{\zeta}{3\gamma - \Omega} B_1 \tag{8a}$$

$$A_2 = -\frac{3\gamma + \Omega}{\zeta} B_2 = -\frac{\zeta}{3\gamma - \Omega} B_2 \tag{8b}$$

which are consistent with the derived dispersion relation (6). We hence chose $\{A_1, B_2\}$ as the independent variables and set $B_1 = -\frac{3\gamma - \Omega}{\zeta} A_1$ and $A_2 = -\frac{3\gamma + \Omega}{\zeta} B_2$.

*Boundary conditions*

The effective exchange fields for edge and bulk spins are different due to the reduced number of nearest neighbors for the edge sites (1 and 2 nearest neighbors for edge spins in nanoribbons with bearded and zigzag edges boundaries respectively). Just like bulk fields, the edge exchange fields are derived from the Heisenberg Hamiltonian and the second order continuum expansion of the edge magnetizations.

The exchange spin waves boundary conditions are derived from the requirement that edge spins satisfy the bulk equations of motion [8 - 11] which yields the effective boundary equations

$$\vec{M}_e^{A/B} \times (\vec{H}_b^{A/B} - \vec{H}_e^{A/B}) = \vec{0} \qquad (9)$$

where $e$ and $b$ stand for edge and boundary respectively. Nanoribbons with bearded or zigzag edges are characterized by opposite spins on the left and right edges. With only one type of spin on the boundaries, equation (9) yields a system of 2 linear equations in the coefficients $A_1$ and $B_2$. For nanoribbons with zigzag edges, the 2 boundary equations yield the matrix equation

$$M_1 \begin{pmatrix} A_1 \\ B_2 \end{pmatrix} = \begin{pmatrix} e^{dq}a_- & e^{-dq}b_+ \\ e^{-dq}b_- & e^{dq}a_+ \end{pmatrix} \begin{pmatrix} A_1 \\ B_2 \end{pmatrix} = 0$$

with $a_\pm = \left(-\frac{3\gamma \pm \Omega}{\zeta}\right)\left(1 + \frac{aq}{\sqrt{3}} + \frac{a^2q^2}{6}\right) + \gamma$, $b_\pm = \left(1 - \frac{aq}{\sqrt{3}} + \frac{a^2q^2}{6}\right) + \left(-\frac{3\gamma \pm \Omega}{\zeta}\right)\gamma$.

For the bearded edges case, the boundary matrix equation changes to

$$M_2 \begin{pmatrix} A_1 \\ B_2 \end{pmatrix} = \begin{pmatrix} e^{dq}c_- & e^{-dq}d_+ \\ e^{-dq}d_- & e^{dq}c_+ \end{pmatrix} \begin{pmatrix} A_1 \\ B_2 \end{pmatrix} = 0$$

with $c_\pm = 2 - \frac{a^2k_x^2}{4} + \frac{aq}{\sqrt{3}} + \frac{a^2q^2}{12} - \frac{2(3\gamma \pm \Omega)}{\zeta}\gamma$, $d_\pm = \left(-\frac{3\gamma \pm \Omega}{t}\right)\left(2 - \frac{a^2k_x^2}{4} - \frac{aq}{\sqrt{3}} + \frac{a^2q^2}{12}\right) + 2\gamma$.

The determinant of the matrices $M_1$ and $M_2$ should vanish as a necessary condition for the existence of non-zero solutions. To ensure consistency in the developed theory, the determinant is calculated keeping only linear and quadratic terms in $k_x$ and $q$. For the nanoribbons with zigzag edges boundaries, the determinant of $M_1$ yields the characteristic boundary equation

$$f_1 = 4\sqrt{3}aq(-1 + \gamma^2) + \{6(-1 + \gamma^2) + a^2[k_x^2 + q^2(-5 + 2\gamma^2)]\}Tanh(2dq) = 0 \qquad (10)$$

For the nanoribbons with bearded edges boundaries, the determinant of $M_2$ yields a different characteristic boundary equation of the form

$$f_2 = 4\sqrt{3}aq(-1 + \gamma^2) + \{12(-1 + \gamma^2) + a^2[k_x^2(5 - 3\gamma^2) + q^2(-4 + \gamma^2)]\}Tanh(2dq) = 0 \qquad (11)$$

With this approach, the boundary conditions are solved simultaneously on both edges. Real solutions for the boundary equations (10) and (11) yield the decay factors for edge exchange spin wave modes in the nanoribbon with zigzag and bearded edges boundaries respectively. Similarly, the imaginary solutions determine the allowed wavevectors for bulk modes in these nanoribbons. For the imaginary solutions, it is useful to substitute $q = ik_y$ ($k_y$ is the wavevector component along the y-direction for propagating modes) in equations (10) and (11). This yields the equivalent equations

$$g_1 = 4\sqrt{3}ak_y(-1 + \gamma^2) + \{6(-1 + \gamma^2) + a^2[k_x^2 + k_y^2(5 - 2\gamma^2)]\}Tan(2dk_y) = 0 \qquad (12)$$

for zigzag edged nanoribbons and

$$g_2 = 4\sqrt{3}ak_y(-1+\gamma^2) + \{12(-1+\gamma^2) + a^2[k_x^2(5-3\gamma^2) - k_y^2(-4+\gamma^2)]\}\text{Tan}(2dk_y) = 0 \quad (13)$$

for nanoribbons with bearded edges boundaries. Equations (12) and (13) do not admit continuous solutions for finite $d$. The allowed wavevector component $k_y$ along the finite direction of the nanoribbon is hence discretized and the number of solutions depends on the width ($2d$) of the nanoribbon.

*Isotropic nanoribbons ($\gamma = 1$)*

For magnetically isotropic nanoribbons, boundary equations (12) and (13) for propagating spin waves reduce to the simple forms $(k_x^2 + 3k_y^2)Tan(2dk_y) = 0$ and $(2k_x^2 + 3k_y^2)\text{Tan}(2dk_y) = 0$ respectively or simply $Tan(2dk_y) = 0$ for both nanoribbons. The discrete solutions are trivial for this case and are given by $k_y = \pm n\pi/2d$ for both types of nanoribbons ($n$ positive integer). Nevertheless, the allowed values of $k_y$ are different for the two nanoribbon types as the possible values for nanoribbon half width $d$ are different in zigzag and bearded edged nanoribbons. This difference becomes particularly relevant in thin nanoribbons (small $d$ values).

In an interval $[k_y = -L, k_y = +L]$ of the long wavelength part of the Brillouin zone, the number of propagating exchange spin waves is hence given by the integer part of the division $\frac{2Ld}{\pi} + 1$. Here, we assume that + and − solutions of $k_y$ belong to the same mode, which follows from equations (5). Again, the number of propagating modes may differ in the 2 different types of nanoribbons due to the difference in $d$.

For edge modes, the boundary equations (10) and (11) reduce in the isotropic case to $(k_x^2 - 3q^2)Tanh(2dq) = 0$ and $(2k_x^2 - 3q^2)\text{Tanh}(2dq) = 0$ respectively. This yields edge exchange spin waves with decay factors $q = \pm\frac{1}{\sqrt{3}}k_x$ and $q = \pm\sqrt{\frac{2}{3}}k_x$ in nanoribbons with zigzag and bearded edges boundaries respectively. The decay factors in this case are independent of the nanoribbons width and have larger absolute values in nanoribbons with bearded edges boundaries. Substituting in the normalized energy dispersion relation (6) yields the linear dispersion relations $\Omega = |k_x|a$ and $\Omega = \frac{1}{\sqrt{2}}|k_x|a$ for edge modes in nanoribbons with zigzag and bearded edges boundaries respectively.

The calculation of the eigenvectors of $M_1$ and $M_2$ yield the amplitudes $A_1$, $A_2$, $B_1$, and $B_2$ for the determined edge and propagating solutions. For the edge modes, the eigenvectors for both matrices are $(1,0)$ and $(0,1)$. This is fundamentally different from the $(1,1)$ and $(-1,1)$ eigenvectors obtained in our previous study for edge spin waves on honeycomb nanoribbons with armchair boundaries [38]. Consequently, unlike nanoribbons with armchair edges boundaries, nanoribbons with zigzag and bearded edges boundaries do not allow edge spin waves propagating with the same $k_x$ (or same direction) on both edges simultaneously. This is a direct consequence of the nontrivial symmetry underlying these nanoribbons. Nanoribbons with bearded or zigzag edges boundaries are characterized by opposite spins on the left and right edges which is different from the armchair edged nanoribbons, where both sublattices are present on the boundaries. For armchair edged nanoribbons, the inversion of the coordinate axis along the finite width constitutes a symmetry operation (an operation that leaves the system invariant). This is not the case for nanoribbons with bearded and zigzag edges boundaries characterized by a more complex symmetry operation. As illustrated in figure 2, the symmetry operation returning the nanoribbon with zigzag edges boundaries to an equivalent state constitutes of y-axis inversion followed by a $180°$ rotation about the y-axis. After the $180°$ rotation about the y-axis, an edge spin wave propagating with $k_x > 0$ at $y = +d$ is transformed to an edge spin wave propagating with $-k_x < 0$ at $y = -d$. Edge spin waves in zigzag edged nanoribbons hence propagate in opposite directions on the left and right edges. The same argument holds for nanoribbons with bearded edges boundaries.

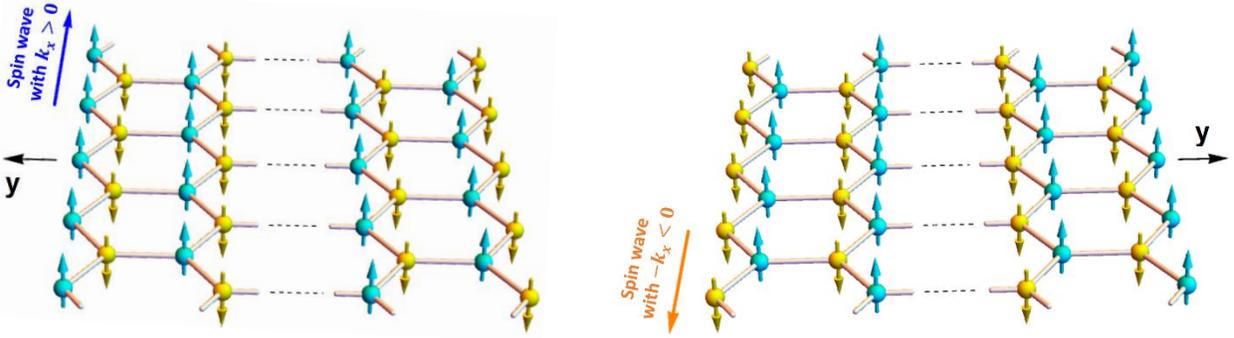

**Figure 2:** Schematic representations illustrating the complex symmetry underlying honeycomb nanoribbons with zigzag edges boundaries (see text for details).

For the numerical applications, we choose $d = \frac{71}{2\sqrt{3}} \approx 20.5$ (the lattice constant $a$ is set to 1) and $d = \frac{70}{2\sqrt{3}} \approx 20.21$ for nanoribbons with zigzag and bearded edges boundaries respectively. The eigenvectors determine the spatial variation of the amplitudes of the edge spin wave across the finite width of the nanoribbon. In figure 3, the normalized $M_x^A$ amplitudes for edge spin waves are plotted along the width of the nanoribbons for $k_x = 0.1$ and $0.3$. The normalized $M_x^B$ amplitudes are out of phase with respect to $M_x^A$ amplitude (from equation (8)) and are not presented here. Compared to zigzag edged nanoribbons, evanescent modes in the nanoribbon with bearded edges boundaries are observed to decay faster as they penetrate into the bulk due to their larger decay factor.

The normalized energies for bulk and edge exchange spin wave modes can be calculated using the dispersion relation presented in equation (6). In the long wavelength part $|k_y| \leq 0.3$ of the Brillouin zone, the boundary condition equation $Tan(2dk_y) = 0$ yields 4 propagating modes for each of the nanoribbons with zigzag and bearded edges boundaries. The discrete wavevectors are $k_y = 0$, $|k_y| = \frac{\sqrt{3}\pi}{71}$, $|k_y| = \frac{2\sqrt{3}\pi}{71}$, and $|k_y| = \frac{3\sqrt{3}\pi}{71}$ for the nanoribbon with zigzag edges boundaries. These are slightly different for the nanoribbon with bearded edges boundaries, with $k_y = 0$, $|k_y| = \frac{\sqrt{3}\pi}{70}$, $|k_y| = \frac{2\sqrt{3}\pi}{70}$, and $|k_y| = \frac{3\sqrt{3}\pi}{70}$. In figure 4, the normalized energy dispersion curves for propagating (blue) and evanescent (red) modes are plotted as a function of the continuous wavevector $k_x$ in the long wavelength portion of the Brillouin zone. The same dispersion curves are plotted in figure 5 as functions of the continuous $k_x$ wavevector for the allowed $(k_y, q)$ values, $k_y$ for (*bulk*) and $q$ for (*evanescent*). The bulk dispersion curves are duplicated at positive and negative values of $k_y$ following the infinite system conventional presentation of the dispersion curves.

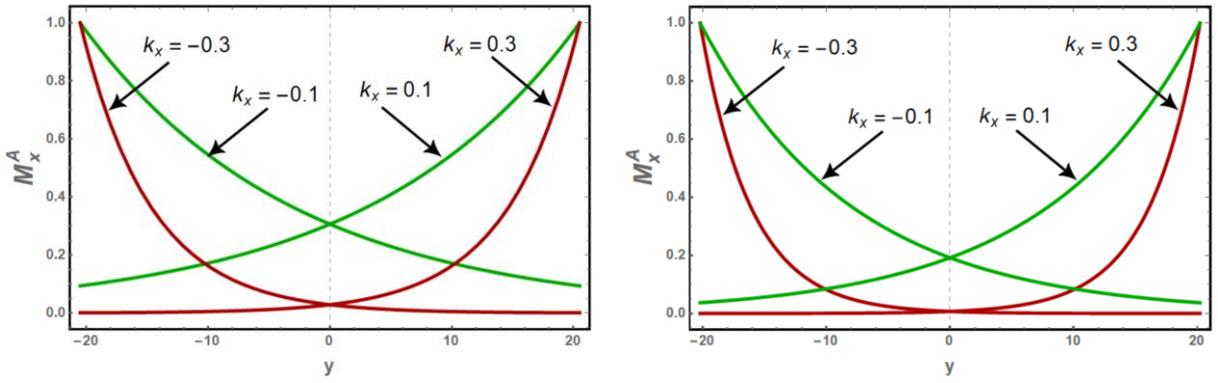

**Figure 3:** The spatial variation of normalized amplitude $M_x^A$ of the nonreciprocal edge spin waves along the finite width of the isotropic nanoribbon with zigzag edges boundaries (left) and bearded edges boundaries (right). The half width is $d = 71/(2\sqrt{3}) \approx 20.5$ and $d = 70/(2\sqrt{3}) \approx 20.21$ for nanoribbons with zigzag and bearded edges boundaries respectively.

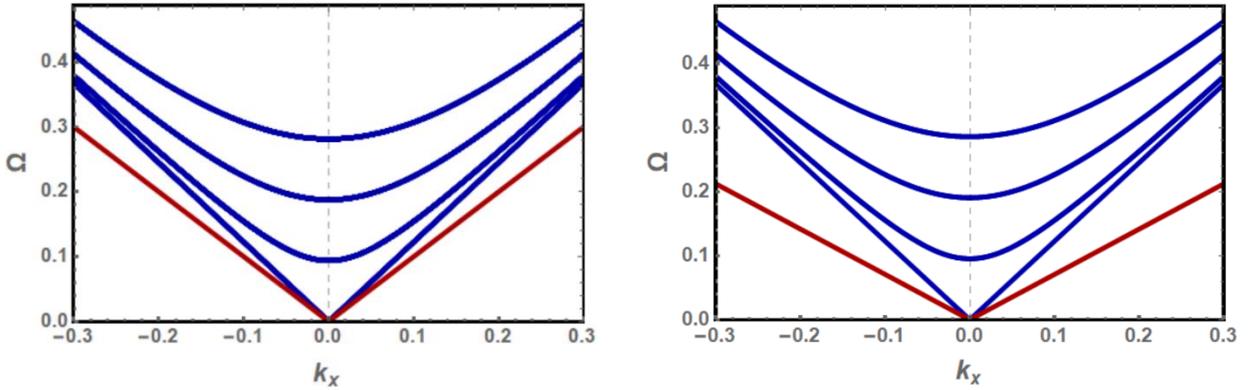

**Figure 4:** Normalized energy dispersion curves as a function of the continuous wavevector $k_x$ for the discretized bulk modes (blue) and the edge mode (red) in the isotropic nanoribbon with zigzag edges boundaries (left) and bearded edges boundaries (right). The half width is $d = 71/(2\sqrt{3})$ and $d = 70/(2\sqrt{3})$ for nanoribbons with zigzag and bearded edges boundaries respectively.

The bulk discrete modes are effectively the same in both nanoribbons and the discretization reduces the Dirac cone in the infinite system to a single linear dispersion curve in the bounded nanoribbons. Another effect of this discretization is the significant energy gaps induced between the allowed bulk spin waves. Compared to our previous results on armchair edged nanoribbons with comparable width [38], the number of allowed bulk wavevectors is reduced which allows the formation of larger energy gaps between modes corresponding to consecutive wavevectors.

The edge modes are Dirac-like modes with energies below the propagating modes energies. They are characterized by a smaller group velocity and are protected by an energy gap. The energy gap between propagating and edge modes is larger for the nanoribbon with bearded edges boundaries. For the zigzag nanoribbon, our results for the edge mode in the isotropic case are identical to previous studies [25, 33] based on the Holstein-Primakov quantum approach and the linear spin wave approximation.

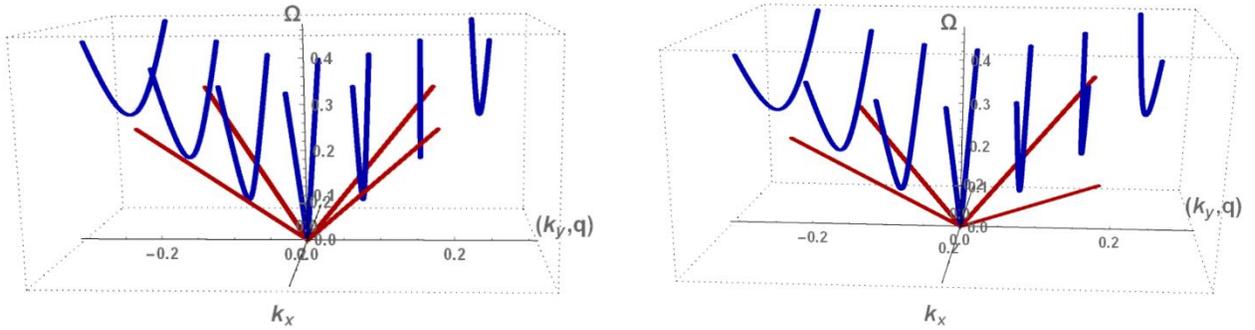

**Figure 5:** Normalized energy dispersion curves for long wavelength bulk (blue) and evanescent (red) exchange spin waves on a magnetically isotropic nanoribbon with zigzag edges boundaries (left) and bearded edges boundaries (right). The half width is $d = 71/(2\sqrt{3})$ and $d = 70/(2\sqrt{3})$ for nanoribbons with zigzag and bearded edges boundaries respectively. The curves are plotted as functions of the continuous $k_x$ wavevector for the allowed $(k_y, q)$ values, $k_y$ for (*bulk*) and $q$ for (*evanescent*).

*Effect of Anisotropy*

We investigate the effect of magnetic anisotropy with $\gamma = 1.01$ and $\gamma = 1.04$ on the bulk and edge exchange spin waves. The nanoribbon width is kept as before. The decay factors solutions $q(k_x)$ for edge modes are determined from the contour plot of equations (12) and (13). The solutions are plotted as a function of $k_x$ in figures (6) and (7) for $\gamma = 1.01$ and $\gamma = 1.04$ respectively.

The slight anisotropies significantly increase the decay factors which are no more in linear dependence on $k_x$. For $\gamma = 1.04$, the decay factors are less dispersive (figure 7) and the spatial variation of the edge spin wave amplitudes along the nanoribbon width become effectively independent of $k_x$ in the long wavelength part. The spatial variations of the edge spin wave amplitudes are plotted in figures (8) and (9) for $\gamma = 1.01$ and $\gamma = 1.04$ respectively. The penetration length for the edge spin waves is reduced significantly due to the slight anisotropies and the evanescent modes are more confined to the edges.

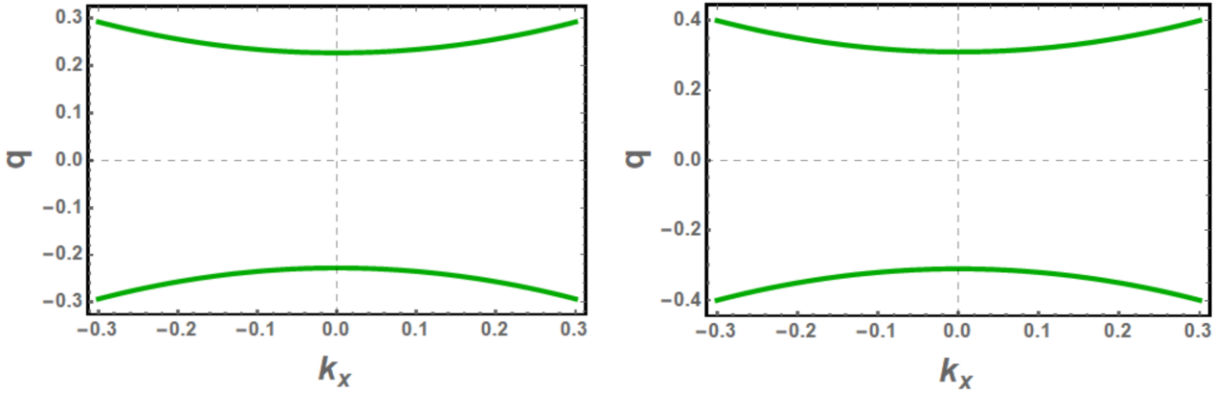

**Figure 6:** The decay factors $q$ as a function of the continuous wavevector $k_x$ for a zigzag edged nanoribbon (left) and bearded edge (right) with slight anisotropy $\gamma = 1.01$. The half width is $d = 71/(2\sqrt{3})$ and $d = 70/(2\sqrt{3})$ for nanoribbons with zigzag and bearded edges boundaries respectively.

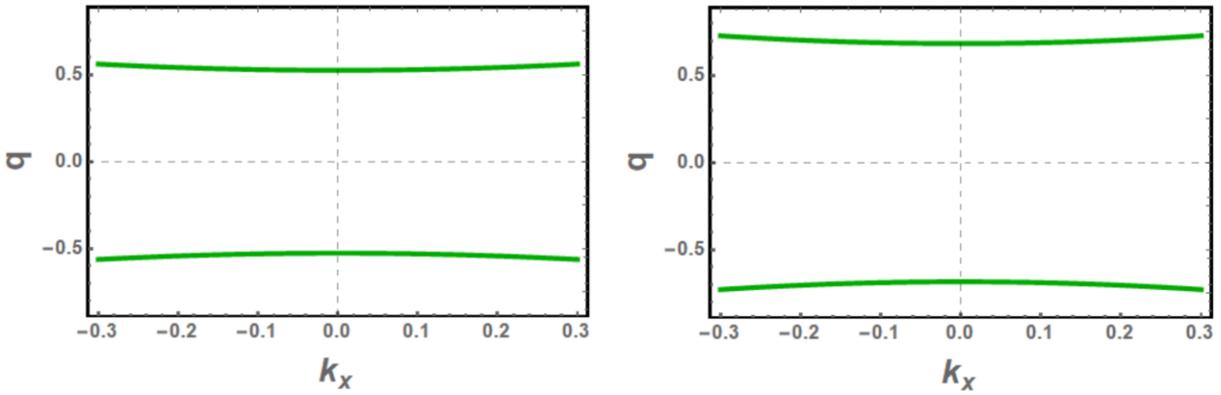

**Figure 7:** Same as figure 6 but with $\gamma = 1.04$.

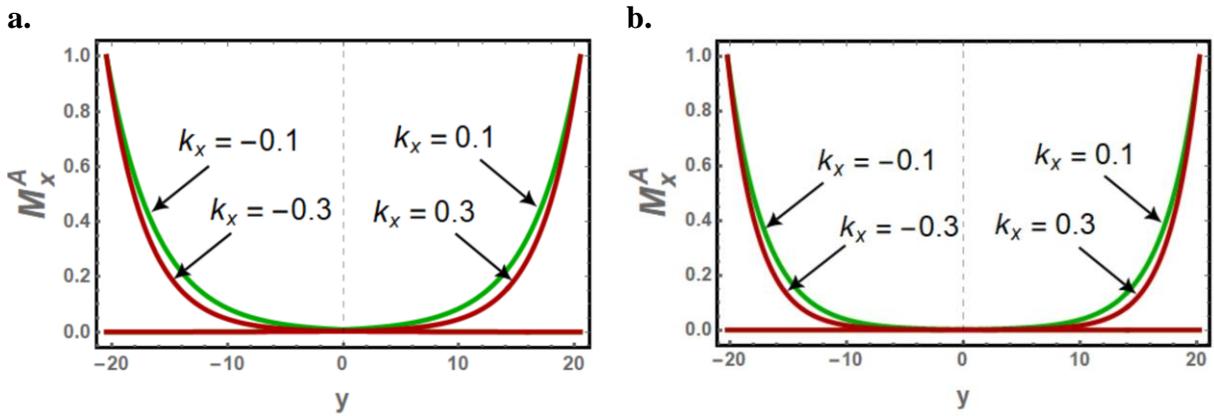

**Figure 8:** Same as figure 3 but for $\gamma = 1.01$.

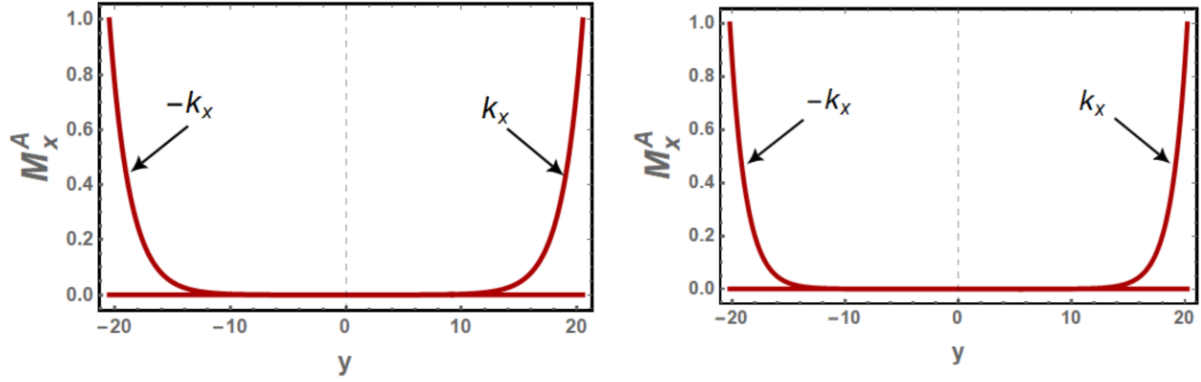

**Figure 9:** Same as figure 3 but for $\gamma = 1.04$.

For the two choices of the anisotropy parameters, the number of bulk modes in the interval $|k_y| \leq 0.3$ is 4 for both nanoribbons. For the nanoribbon with zigzag edges boundaries, the allowed $k_y$ wavevectors are $k_y = 0$, $k_y \approx \pm 0.075$, $k_y \approx \pm 0.151$, and $k_y \approx \pm 0.227$ for $\gamma = 1.01$ and $k_y = 0$, $k_y \approx \pm 0.075$, $k_y \approx \pm 0.149$, and $k_y \approx \pm 0.225$ for $\gamma = 1.04$. As the anisotropy increases, the allowed wavevectors shift slightly toward the origin but the significant spacing between consecutive wavevectors is barely affected. We note that this shift of $k_y$ towards the origin in zigzag edged nanoribbons with large anisotropy can allow additional propagating modes in the $|k_y| < 0.3$ interval. For example, the number of modes in the zigzag edged nanoribbon with $\gamma = 1.1$ becomes 5, with an additional mode at $k_y \approx \pm 0.299$.

For the nanoribbon with bearded edges boundaries, the allowed $k_y$ wavevectors are $k_y = 0$, $k_y \approx \pm 0.0767$, $k_y \approx \pm 0.1537$, and $k_y \approx \pm 0.2312$ for $\gamma = 1.01$ and $k_y = 0$, $k_y \approx \pm 0.0766$, $k_y \approx \pm 0.1533$, and $k_y \approx \pm 0.2301$ for $\gamma = 1.04$. The allowed wavevectors also shift toward the origin and this shift is smaller for the nanoribbon with bearded edges boundaries. Also, the allowed wavevectors in the two types of nanoribbons are very close in the relatively wide nanoribbons under study.

The normalized energy dispersion curves for propagating (blue) and evanescent (red) modes are plotted as a function of the continuous wavevector $k_x$ in figures (10) and (11) for $\gamma = 1.01$ and $\gamma = 1.04$ respectively. The same dispersion curves are plotted in figures (12) and (13) as functions of $k_x$ wavevector for the allowed $(k_y, q)$ values, $k_y$ for bulk and $q$ for evanescent.

The normalized energies for the discrete bulk spin waves are effectively the same for the zigzag and bearded edges nanoribbons. Although the spacing between consecutive discrete $k_y$ solutions are preserved, the energy gap between the bulk modes energies is reduced significantly with increasing anisotropy. Moreover, the dispersion of the bulk and edge modes energies decreases with anisotropy and the spin waves are characterized by slower group velocities. The density of states, however, is significantly enhanced.

A large energy gap is induced by anisotropy between the energies of propagating and edge spin waves in both nanoribbons. The gap is observed to increase with anisotropy. This large gap allows the separate excitation of long wave-length bulk and edge exchange spin waves. The edge mode energy is significantly lower in the nanoribbon with bearded edges boundaries due to the larger decay factor. An interesting observation is that the edge mode energy in this nanoribbon is lower in the $\gamma = 1.04$ case compared to the $\gamma = 1.01$.

As mentioned previously, our results for the zigzag edged nanoribbon differ significantly from previous studies due to the more general form adapted for the magnetization dynamics and the consequent treatment of the resulting boundary conditions.

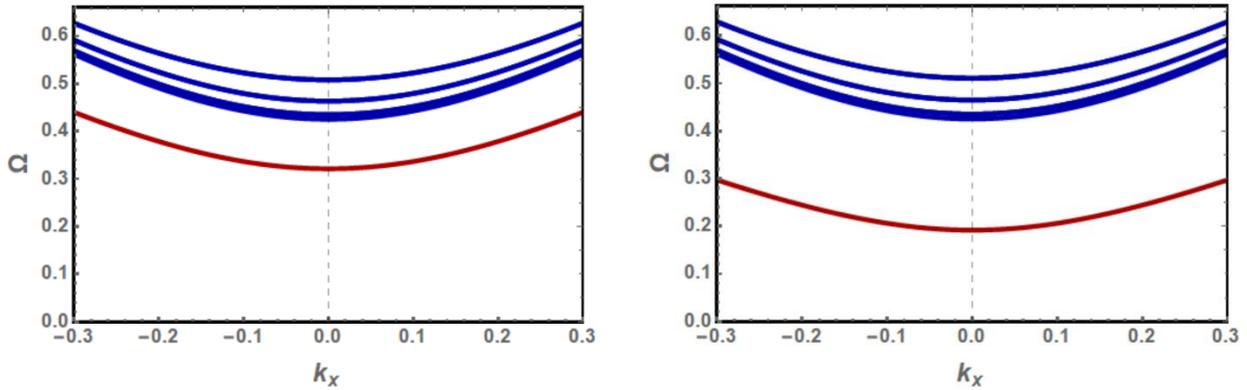

**Figure 10:** Same as figure 4 but for $\gamma = 1.01$.

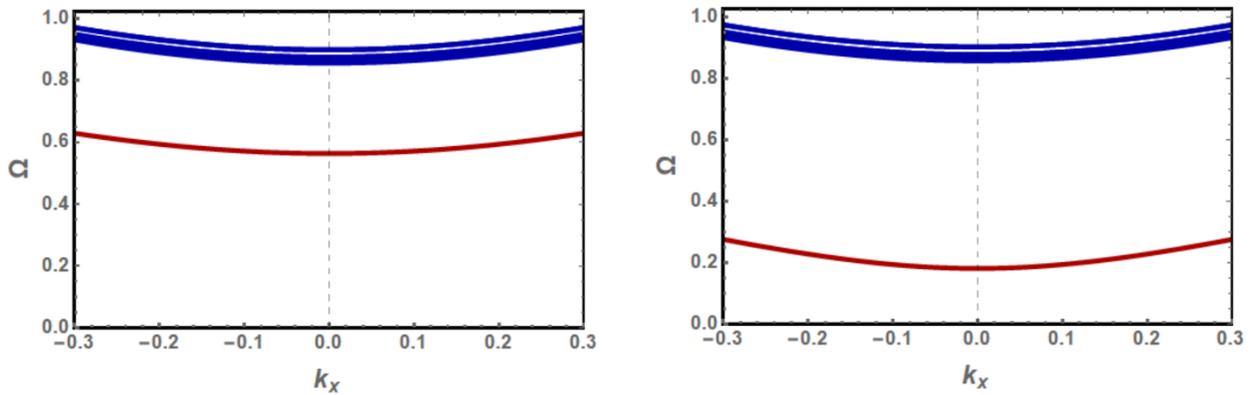

**Figure 11:** Same as figure 4 but for $\gamma = 1.04$.

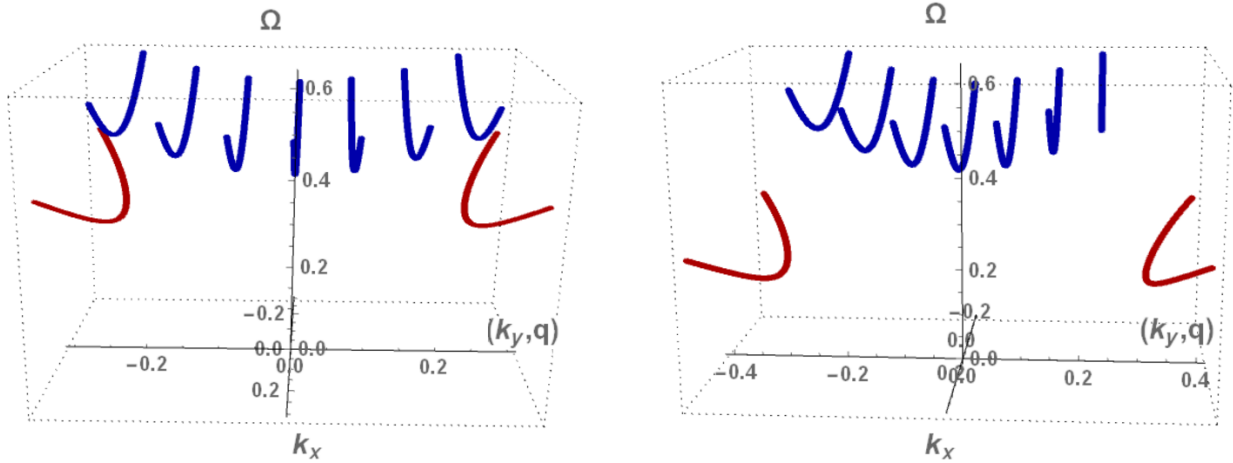

**Figure 12:** Same as figure 5 but for $\gamma = 1.01$.

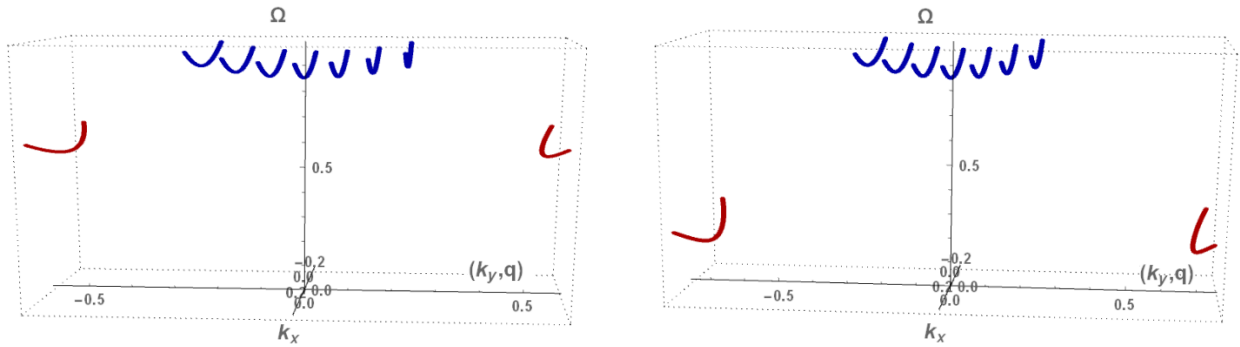

**Figure 13:** Same as figure 5 but for $\gamma = 1.01$..

## *Conclusion*

A classical field theory approach is developed to calculate the long wavelength exchange spin wave modes on honeycomb nanoribbons with zigzag and bearded edges boundaries. Appropriate boundary conditions, which takes into account the finite width of the nanoribbon, are derived from the requirement that edge spins satisfy the bulk Bloch equations of motion. The boundary equations are solved simultaneously on both edges to determine the allowed decay factors and discrete wavevector component for edge and propagating modes respectively. Solving the bulk equations of motion simultaneously with the edge boundary conditions equations determines the edge and bulk spin waves.

Our theoretical study predicts that nanoribbons with zigzag and bearded edges boundaries forbid the simultaneous propagation of edge spin waves in the same direction on opposite edges. The edge spin waves hence propagate in opposite directions on the left and right edges of these nanoribbons, in analogy with nonreciprocal surface spin waves in magnetic thin films. The existence of these edge spin waves, like nonreciprocal surface modes, is directly related to the nontrivial symmetry underlying the nanoribbons. Such edge modes, for example, cannot exist in nanoribbons with armchair boundaries. Despite their opposite propagation directions, the edge modes have the same propagation frequency in the absence of an external magnetic field in our study.

In the absence of magnetic anisotropy, the dispersion relation for edge modes is linear for both nanoribbons. The energy gap between edge modes and propagating modes is found to be larger in nanoribbons with bearded edges boundaries. This follows from the larger decay factors characterizing the edge modes in nanoribbons with bearded edges boundaries. Bearded edge spin waves are hence more confined in to the edges compared to zigzag edge spin waves.

Magnetic anisotropy is found to significantly increase the energy gap between edge and propagating modes in both nanoribbon types. The large energy gap allows for separate excitation of edge and bulk modes which are no more overlapping for $\gamma = 1.04$. Moreover, the magnetic anisotropy increases the edge modes decay factors, reduces their group velocities, and increases their density of states. The energy gap and decay factors stay larger for nanoribbons with bearded edges boundaries.

Our results for anisotropic nanoribbons with zigzag edges boundaries differ significantly from previous studies based on quantum approaches for two main reasons. First, we have used a more general form for the magnetization dynamics presented in equations (5). Second, the derived boundary equations based on this form of the solution are solved simultaneously on both edges.

The bulk modes are found to be discretized due to the finite width of the nanoribbon and the Dirac cone in the infinite honeycomb lattice is reduced to a single Dirac mode. For the relatively wide nanoribbons ($d \approx 20\, a$) chosen for numerical applications, the allowed discrete wavevectors along the finite width of different nanoribbon types are very close. The bulk energy spectrum is hence effectively identical for both nanoribbons. This is not the case for thin nanoribbons.

Significant energy gaps are found between the bulk discretized modes in the isotropic nanoribbons. For anisotropic nanoribbons, the energy gaps between propagating modes and the group velocities are reduced while the density of states increases. The discrete wavevector component solutions are slightly shifted towards the origin for anisotropic nanoribbons which might allow additional propagating modes for large values of the anisotropy parameter.

Similar to nonreciprocal surface spin waves, the asymmetric edge spin waves predicted on nanoribbons with zigzag and bearded edges boundaries are expected to be very interesting for magnonics applications. Edge spin wave excitations in 2D Dirac materials is a relatively recent field compared to surface spin waves and theoretical studies are indispensable to support or even guide experimental work on these exotic materials. In this context, further development of the classical field theory beyond the scope of the present work (restricted to exchange and anisotropy interactions) is indeed necessary.